# THE LABORATORY COMPLEX FOR THE CALIBRATION OF PHOTOMETERS USING THE OPTICAL METHOD FOR DETERMINATION OF THE WATER VAPOR CONTENT IN THE EARTH ATMOSPHERE

Galkin V. D.[1)], Sal'nikov I. B.[1)], Nikanorova I. N.[1)], Leiterer U.[2)], Naebert T.[2)], Alekseeva G. A.[1)], Novikov V. V.[1)], Ilyin G. N .[1)], and Pakhomov V. P.[1)]

1) The Central (Pulkovo) Astronomical Observatory, Russian Academy of Sciences, Russia
2) Deutscher Wetterdienst, Meteorologisches Observatorium Lindenberg, Germany

**Abstract**

*We describe the laboratory complex for the calibration of photometers that are used in weather service to measure the water vapor content in the Earth atmosphere. The complex was built up in Pulkovo Observatory and developed within the framework of collaboration between Pulkovo Observatory and Lindenberg Meteorological Observatory (Meteorologisches Observatorium Lindenberg – Richard-Aßmann-Observatorium, Lindenberg, Germany). It is used to obtain calibration dependences for individual devices, and also to develop and compare various methods of construction of calibration dependences. These techniques are based on direct calibration of the photometers, on the use of spectral laboratory transmission functions for water vapor, on calculation methods using spectroscopical databases for individual lines. We hope that when the parameters of the equipment are taken into account in detail and new results for the absorptive power of water vapor are used, the accuracy of determination of the water vapor content in the atmosphere of 1-2% may be attained.*

## Introduction

Climatic variations on the Earth, the global warming or cooling, accurate short- and long-term weather forecasting are general human problems rather than problems of any local field of science. Water vapor plays a key role in these issues. Its ability to adopt various aggregate states, to accumulate vast amounts of energy and transfer it over large distances, to give it out to the ocean and resume it later – all these unique features of the water vapor are closely related to physical processes in the atmosphere. The accuracy of the measurements of the water vapor content in the atmosphere largely specifies the reliability of weather forecasting and the possibility to reveal long-term trends in climate variations. Currently, different methods for the measurements of the atmospheric water vapor content are developed, such as space- and ground-based microwave methods, global coordinate determination systems (GPS), lidar measurements and, of course, networks of atmosphere radiosonde stations. Among these techniques, a particular position belongs to the optical method, in which the absorption of light by the water vapor is measured in spectra of the Sun or stars and, using the calibration dependence of the absorption on the amount of the water vapor along the line of view, its content in the Earth atmosphere is determined. The optical method is the oldest of all methods of determination of atmospheric water vapor content [1]. It is commonly implemented with the use of solar photometers of different design and reaches the accuracy of determination of the water vapor content in the atmosphere of the order of 10% [2-5]. At Pulkovo Observatory, the optical method was substantially developed and for the first time applied for the determination of the night-time atmospheric water vapor content [6, 7] the on the base of distributions of energy in the spectra of stars taken from the homogeneous Pulkovo spectrophotometric catalog [8]. Later on, owing to our long-term fruitful collaboration with the Lindenberg Meteorological Observatory (Germany), for the first time 24-hour optical monitoring of atmospheric water vapor content was implemented in Lindenberg, with the use of solar and stellar photometers, and the accuracy of determination of the atmospheric water vapor content reached 3-5% [9-14].

In differential photometrical observations, accuracy of the order of 0.5% is not exceptional. If this accuracy could be maintained through the processing of the observations up to the final result – the amount of the water vapor along the line of view – the accuracy of determination of the atmospheric water vapor content could reach 1-2%. With this accuracy, and taking into account the fact that the optical method is totally stand-alone (that is, it does not require calibration with the use of outer data) the results obtained with this technique may be used to calibrate and verify other methods.

Two problems arise in the process of determination of the atmospheric water vapor content with the optical method. First, from the observed extinction of light in the atmosphere, its part specified by the absorption by water vapor should be discriminated, and the relation between the observed absorption and the amount of water vapor along the line of view should be obtained. The latter relation may be derived from the known parameters of the filters and the absorption in the atmosphere calculated with the use of the available HITRAN spectroscopic database [15] and various models of absorption averaging, like LOWTRAN, MODTRAN etc. [2-5]. Currently, this is the most widely spread method of calibration. Another calibration technique, which is rarely used because of its labor content, but which is more reliable metrologically, is the direct laboratory calibration of the dependence of the absorption on the number of absorbing molecules. In the latter case, under controlled laboratory conditions, the number of absorbing molecules equivalent to their number in the atmosphere, with temperature and pressure values typical for the atmosphere, should be provided. This may be attained with optical paths of the order of several kilometers, which is by no means simple a problem under laboratory conditions. We used the unique Pulkovo VKM-100 multipass vacuum cell with the base length 96.7 m, which provides the optical path of several kilometers with the use of a set of mirrors for multiple light reflections. In addition to the variation of the length of the optical path by the number of passages, the concentration of the water vapor in the cell may also be adjusted by variation of relative humidity of the air within the cell. Thereby, various absorption values may be reproduced for typical values of humidity and pressure in the atmosphere. With regard to temperature, we have thus far been able to provide a uniform temperature distribution along the tunnel only for two temperature values: for $t = +13^{o}$ C with the steam heating off and for $t = 20^{o}$ C with the heating on. Note that, in order to attain the accuracy of determination of the amount of water vapor 1-2%, not only the calibration, but also stability of the entire photometric system and its reaction to variation of external parameters should be taken into account. We hope that our combined approach, which includes the study and verification of our instrumental system both under laboratory and outdoor conditions, will make it possible to reach the above accuracy. A clue for this is the construction of a laboratory complex for the calibration and testing of the equipment for accurate measurements of the water vapor content in the Earth atmosphere.

**The optical method for determination of the amount of water vapor along the line of view.**

The optical method for determination of the water vapor content suggests consecutive simultaneous photometrical measurements at the wavelengths within an absorption band of the water vapor and beyond it. Figure 1 presents the filter wavelengths relative to the 930 nm water vapor absorption band, for the Pulkovo PZF-94 photometer.

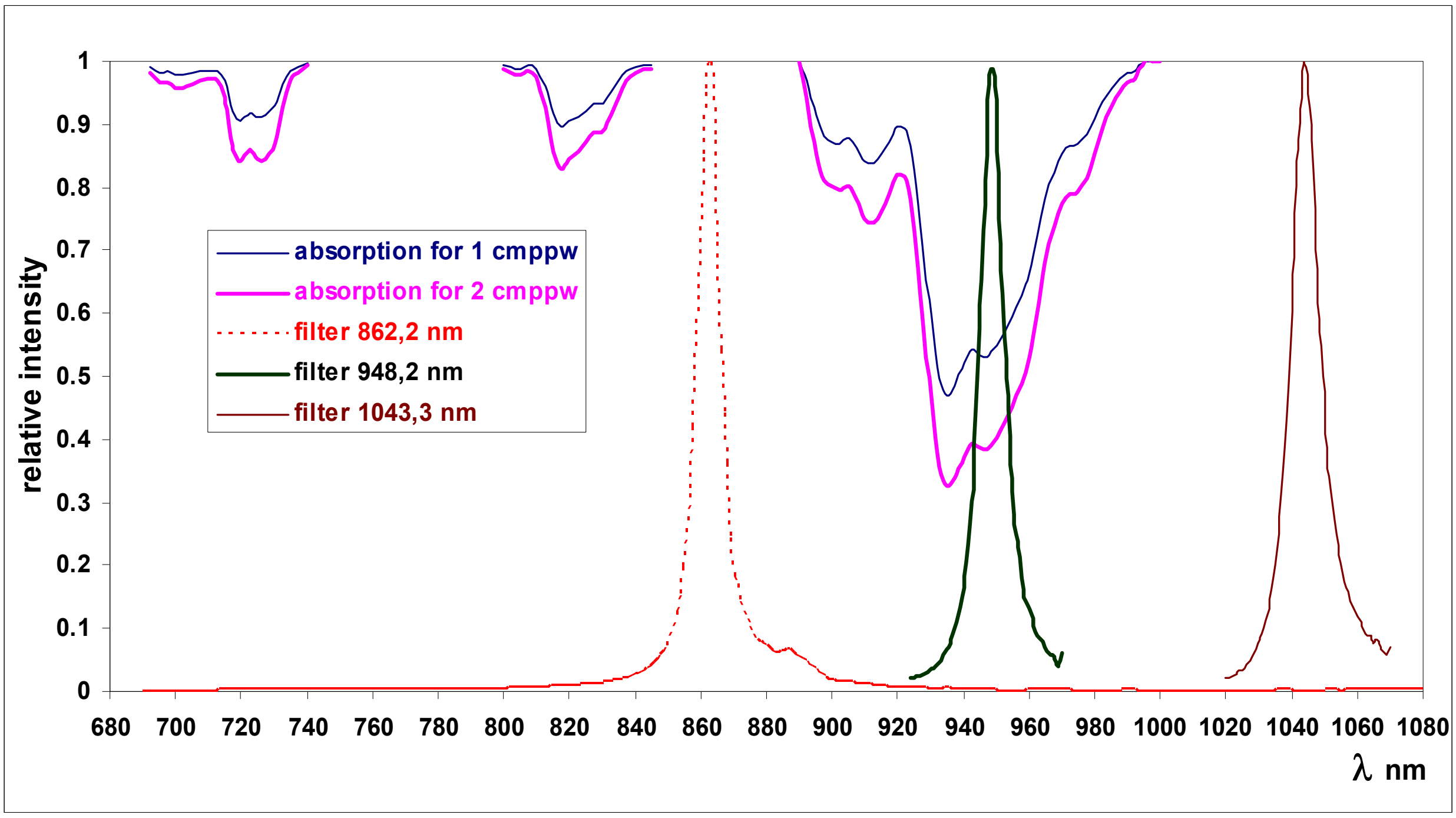

**Fig. 1**. The position of the filter wavelengths of the PZF-94 photometer relative to the water vapor band.

In spectrophotometrical observations, the position of a spectral interval used to measure a signal is determined by the position of the rectangular exit slit of the instrument, and may be optimally selected both inside the band and out of it. The difference of the measurements in the portions of the spectrum located within the absorption band and beyond it, corrected for aerosol extinction, characterizes the absorptive power of water vapor at the given wavelength and depends on the number of absorbing molecules of water vapor, the pressure, and the temperature of the absorbing medium. In addition to that, the observed absorption in the band depends on the parameters of the used filter (its width and the shape of the transmittance curve). Of all these dependences, the most substantial is that on the number of absorbing molecules. In the Earth atmosphere, the content of water vapor varies from 0.1 to 6 cm of precipitated water. Taking into account that in the observations the air mass may reach 10, we may conclude that we should study the absorption dependence in the water vapor bands for the content of water vapor from 0.1 to 60 cm of precipitated water along the line of view. Precisely for this interval of the variation of the water vapor content, the laboratory absorption data should be calculated or obtained. In a given point of the observations, the effective pressure and temperature for atmospheric water vapor do not deviate from their average values more than by 5%. In the optical method, a filter or the exit slit of the spectrophotometer average the absorption in spectral lines within the detected wavelength interval. According to the statistical model, the absorption caused by a set of spectral lines may be represented by the expression [16]:

$$A=1-T=1-\exp\{-\Sigma W_i\}, \quad (1)$$

where **A** is the absorption produced by the set of spectral lines, **T** – transmittance, $\mathbf{W_i}$ – the equivalent width of an individual spectral line. The expression (1) makes it possible to calculate the absorption produced by multiple lines as a function of the amount of water vapor along the line of view, its pressure and temperature, provided the spectroscopical parameters of individual lines, which are necessary to construct the growth curve for each line, are known. Although the expression (1) makes it possible to calculate absorption, it does not yield analytical dependence of the absorption on the number of absorbing molecules of the water vapor **ω**, on its pressure **P,** and the temperature **T**. An appropriate empirical approximation to this dependence, at least to that on the physical parameters **ω** and **P**, is given by the power function:

$$T = \exp\{-\beta \cdot \omega^{\mu} \cdot P^{n}\}, \quad (2)$$

where **β**, **ω**, **n** are empirical parameters. Transforming (2) to stellar magnitudes, we obtain:

$$[m\text{-}mo](\omega) = c \cdot \omega^{\mu}, \quad (3)$$

where **ω** is the amount of water vapor measured in centimeters of precipitated water, **c** the empirical parameter, which corresponds to absorption in stellar magnitudes for 1 cm of the water vapor along the line of view and depends on the effective pressure of the water vapor as follows:

$$c = (-2{,}5 * \log(e)) * P^{n}, \quad (4)$$

The empirical parameters may be obtained either from the approximation of the absorption either calculated for the atmosphere conditions with the use of such models as LOWTRAN, MODTRAN, HITRAN [2-5], or obtained experimentally, under laboratory conditions or in on-sky mode.

**The Pulkovo VKM-100 vacuum multipass cell.**

In laboratory, we can derive the absorption from the comparison of photometrical data obtained for variable amount of water vapor along the line of view with those obtained in the absence of the water vapor (in the evacuated cell). The Pulkovo VKM-100 multipass vacuum cell (Fig. 2a) makes it possible to obtain different numbers of water vapor molecules for a specified pressure, varying the number of passages of light in the cell.

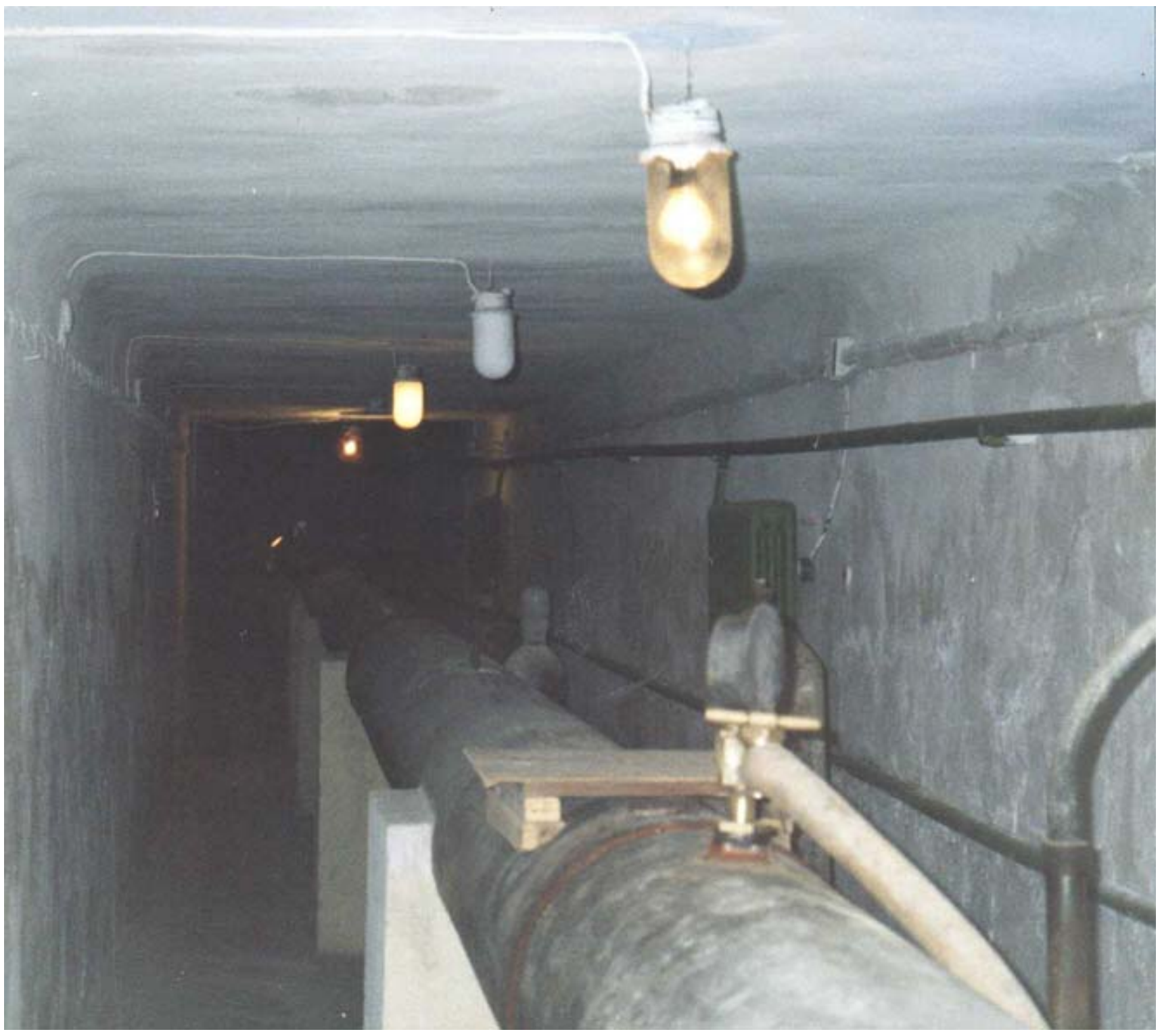

**Fig. 2a**. The VKM-100 cell: the general view after reconstruction.

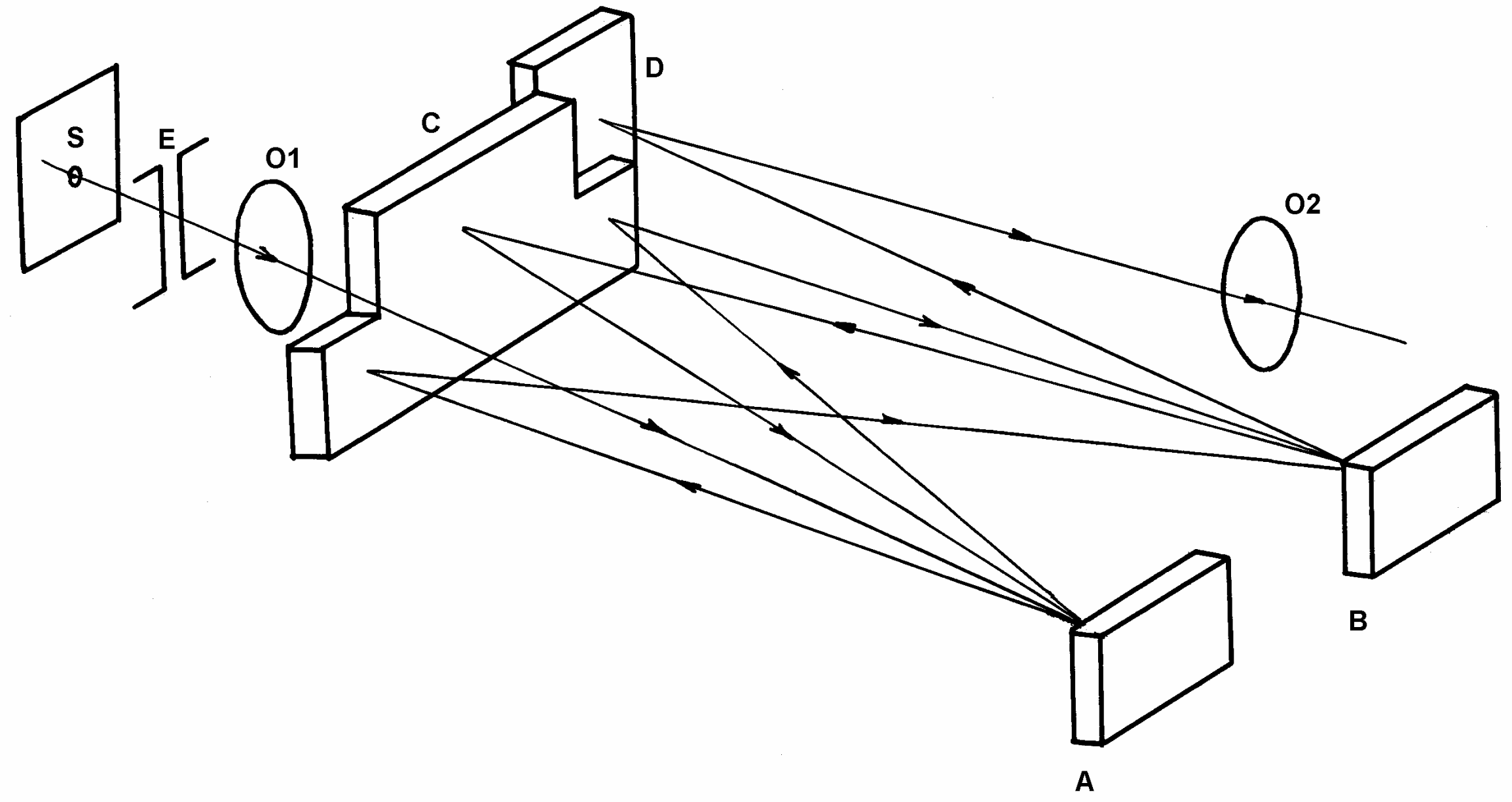

**Fig. 2b**. The general optical schematic diagram of the VKM-100 cell.

Figure 2b presents the general optical schematic diagram of the cell, composed according to White's scheme [17]. Spherical mirrors **A**, **B**, **C,** with the radius of curvature 96.5 m are mounted so that the mirrors **A** and **B** form a sequence of images of the entrance slit on the mirror **C**. The mirror **C** reflects the mirror **A** onto **B,** and vice versa. The entrance lens **O1** mounted in the plane of the entrance slit **E** forms the image of the light source, restricted by the diaphragm **S**, on the mirror **A**. The diaphragm **S** restricts the light beam within the solid angle of the mirror **A**, thereby eliminating superfluous scattered light in the cell. The number of light passages varies due to the variation of the relative position of the optical axes of the mirrors A and B and, hence, to the variation of the number of images on the mirror C. As we can see in Fig.1a, the mirrors A and B should be adjusted so that in the upper row of images formed on the mirror C, some odd number of images is formed; given that, the last (even) image will be placed on the exit slit. In addition to White's scheme, instead of the exit slit, the mirror D is introduced, which reflects the mirror B onto the exit lens O2. Thereby, the system of mirrors A, B, and C, makes it possible to obtain multiple passages of light, starting with the minimum number of passages, equal to 4, and then increasing it by an integer factor. Thus, the images of the entrance slit appear in the exit window of the cell (behind the lens O2) after the number of passages equal to 5=4+1, 9=8+1, 13=12+1, 17=16+1, etc. The maximum number of passages is restricted by the number of images of the entrance slit which can be placed along the mirror C (for the VKM-100 cell, this number reaches a hundred images, which corresponds to the path length of 40 km). In practice, however, the maximum number of passages is substantially lower due to light losses on reflection, which vary as $r^N$, where r is the reflection index, and N the number of reflections.

Figure 3 presents the variation of the intensity of the signal in stellar magnitudes for the filter centered on 672 nm as a function of the path length (the number of reflections). On the path length of 4100 m, the signal decreases by 6 stellar magnitudes (by the factor of 250), according to the reflection index of mirrors ~ 89 % (aluminum covering).

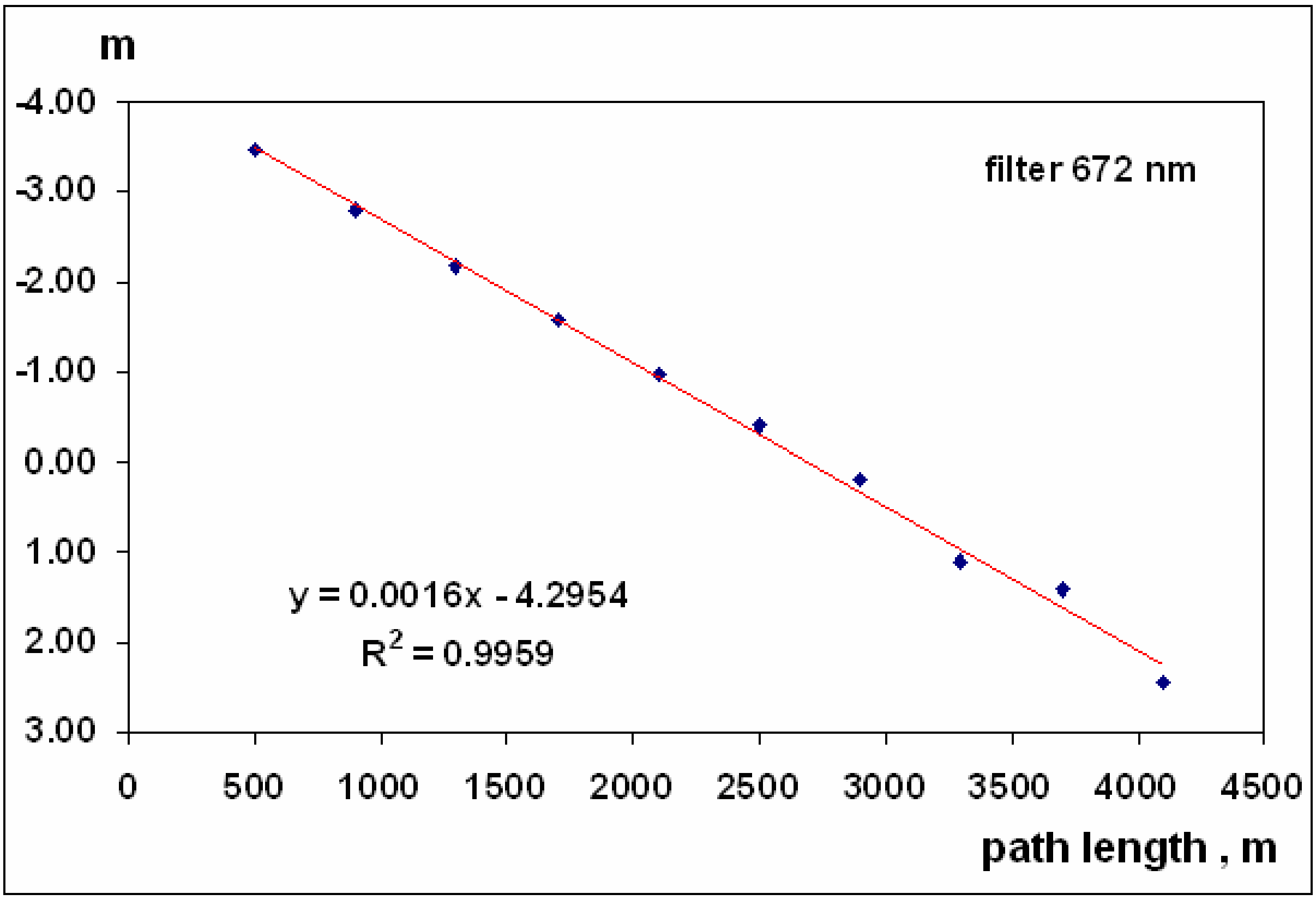

**Fig. 3**. The intensity of light that passes through the cell, as a function of the path length.

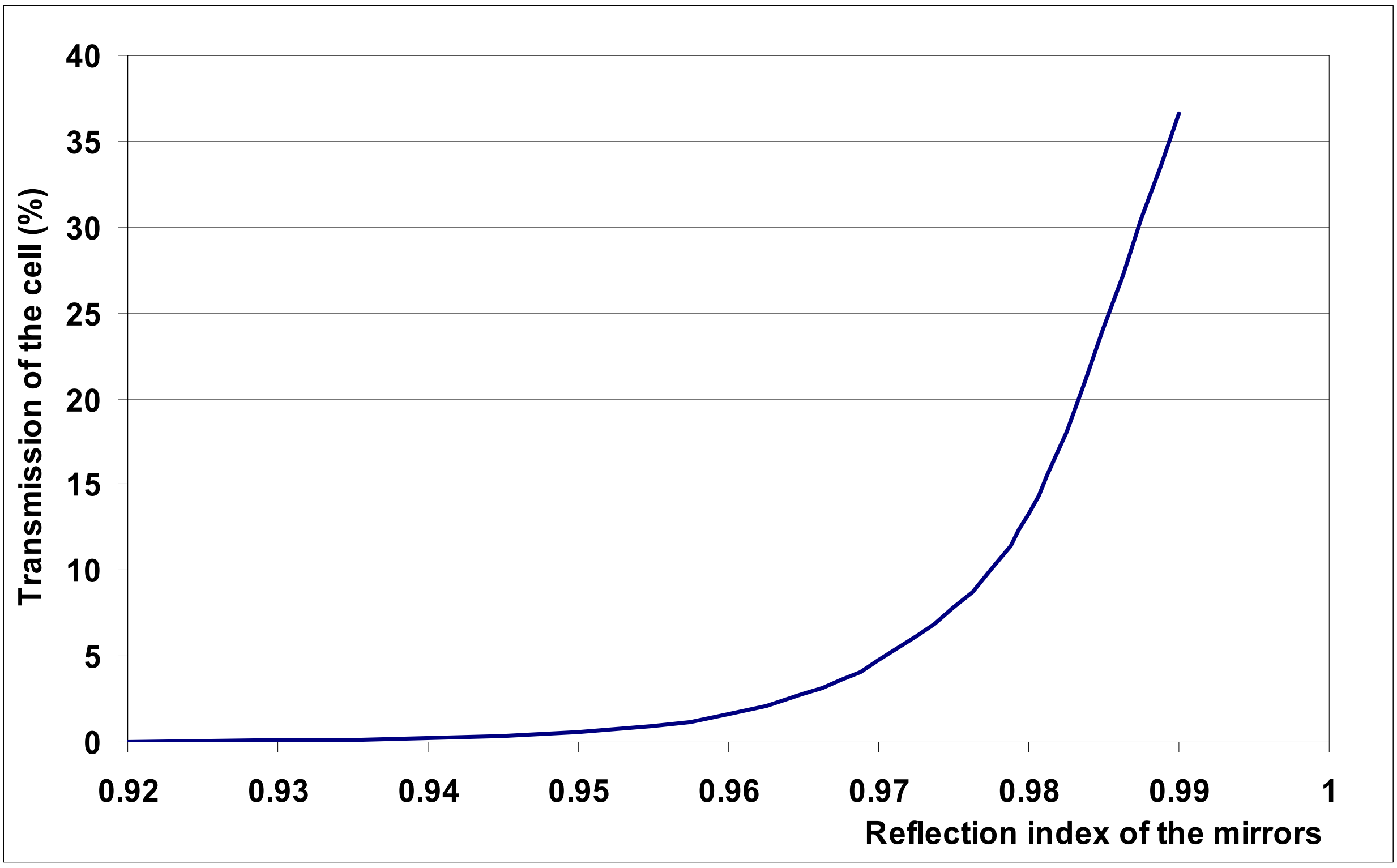

**Fig. 4**. Transmission of the cell for the path length of 10 km as a function of the reflection index of the mirrors.

Figure 4 presents the intensity of the signal (in percent) as a function of the reflection index of the mirrors for the path length 10 000 m (100 reflections). The Figure indicates that a plausible level of the signal may be reached with the reflection index 95-96%, which corresponds to silver-coated mirrors. Another reason limiting the maximum distance that light can pass in the cell is the diffusion of the entrance slit image with the increase in the number of reflections. This is due to insufficient quality of the mirror surfaces caused by difficulties with the testing of the curvature radius for mirrors with such small curvature.

Vitally important is the temperature distribution along the cell. The temperature is naturally homogeneous in summer time, when inhomogeneous local heat sources are absent. The tunnel walls adopt the temperature of the surrounding soil, and along the entire length of the cell the temperature within the interval 11-13$^o$ C is established; the variations of the temperature of the surrounding air result only in variations of the temperature in the cell within this interval. In the heating season, temperature gradient is formed along the tunnel: in its southern end, under the building, the temperature around 20$^o$ C is established, while in the northern end it depends on the outer air temperature, and varies between +10 and -5$^o$ C. To eliminate the temperature gradient, we undertook reconstruction of the tunnel, as a result of which, as it is seen in Fig.5, the temperature levelled off substantially. The temperature is tested with thermometers disposed in the tunnel at the distance of 10 meters one from another, and also with temperature sensors inside the cell.

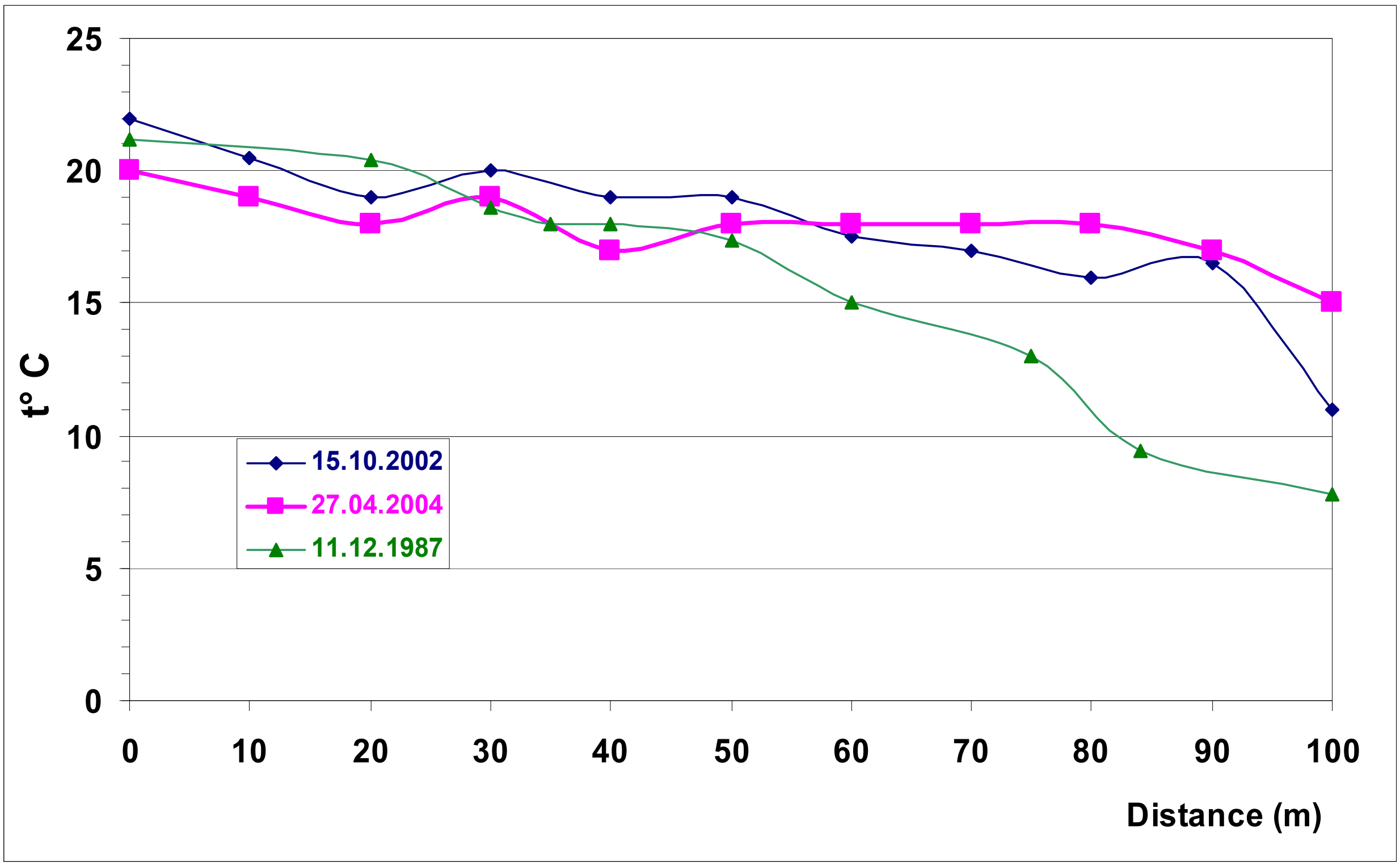

**Fig. 5**. The air temperature distribution along the VKM-100 cell.

Since we aim at studying water vapor under the conditions of the free atmosphere, the cell is filled with air with the natural water vapor content, or with addition of liquid water. The pressure in the cell is established either by evacuation of the cell volume, or by introduction of extra air. To measure the pressure, standardized and technical vacuum gauges and manometers are used.

The amount of water vapor in the cell is measured with humidity sensors mounted at the edges and in the middle of the cell. In order to test the uniformity of the water vapor distribution in the volume of the cell, the observations of the 6943,843 Ǻ water vapor line are also carried out with the ASP-12 high-resolution vacuum diffraction spectrograph (Fig. 6) [18], optically connected with the VKM-100.

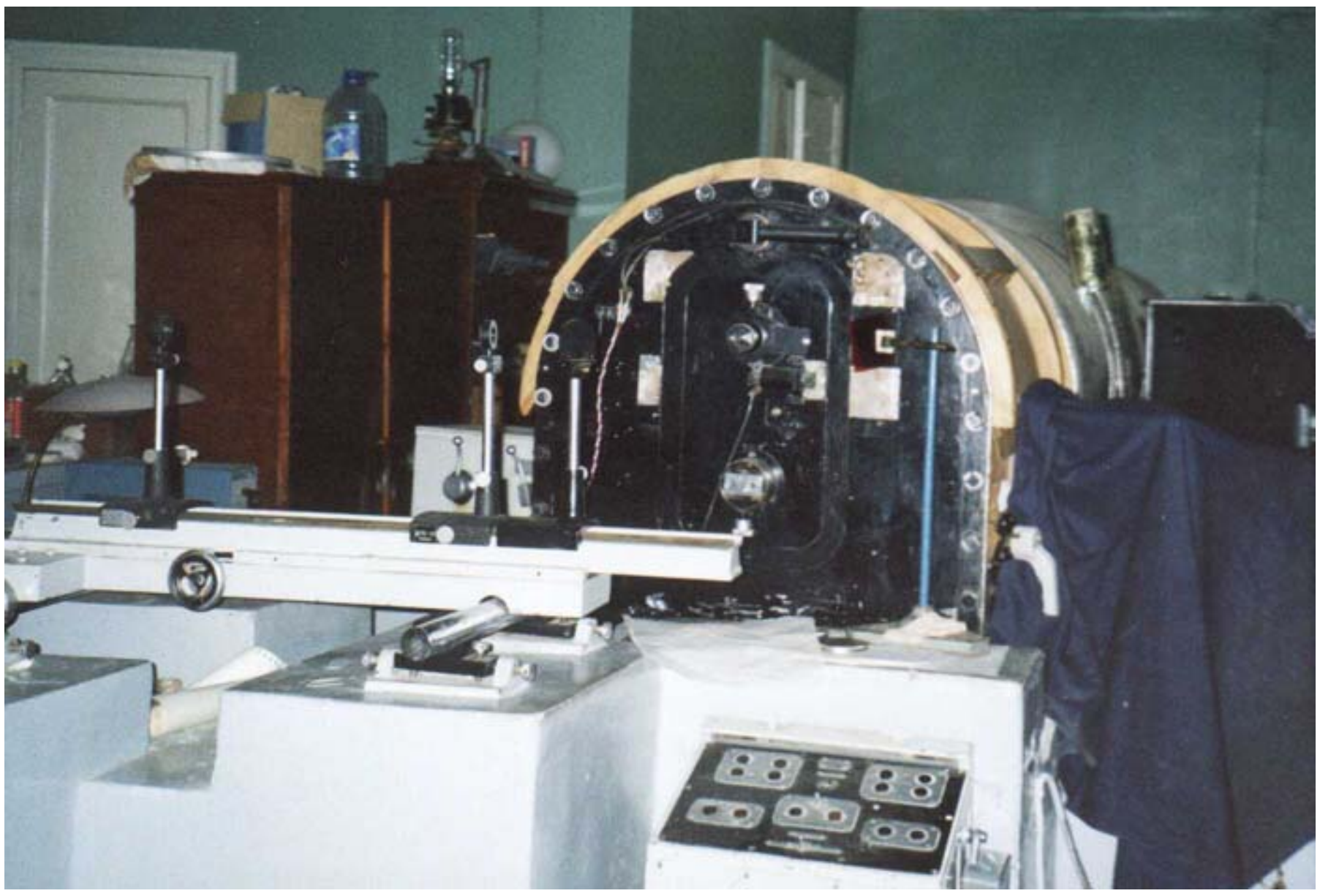

**Fig. 6**. The general view of the ASP-12 vacuum diffraction spectrograph.

**The determination of the water vapor content in the cell.**

The determination of the water vapor content in the cell is a particularly complicated and cumbersome process. The sensors of humidity and relative humidity require permanent testing of their reliability. In addition to that, as the measurements are essentially local, extension of their results on the total volume of the cell is to some extent arbitrary, and the reliability of this operation is difficult to estimate. The pressure of saturated water vapor is specified by the temperature and varies by 6% with a $1^{o}$ C temperature variation. Adoption of an average temperature for the cell in total may result in the introduction of additional error in the calculation of the pressure of saturated vapor and, consequently, in the calculation of the absolute humidity in the cell. To measure the latter parameter, we used sorption-capacitive relative humidity sensors and a microprocessor-based gauge for relative humidity and temperature, which derives the absolute humidity from its measurements. The sensors function on the basis of the dependence of dielectric permeability of polymer sorbent film on the amount of the absorbed moisture. They should be periodically calibrated with reference to standards of the relative humidity of the air. However, the manufacturing company carries out their calibration only under the atmospheric pressure, which does not guarantee invariability of the calibration for other values of pressure. We tested the calibration of our sensors and compared their readings for various pressure values with the data obtained with the bench tester built for similar purposes in the Lindenberg Meteorological Observatory. For the calibration, we used a standard saline generator of moisturized air, the action of which is based on the fact that saturated saline solutions under a constant temperature form and maintain a constant relative humidity of air in closed space. The use of the MERCK (molecular moisture absorbent), LiCl, MgCl, NaCl solutions and double $H_2O$ distillate made it possible to establish the relative humidity values of 0, 11.3, 33.1, 75.5, and 100%, respectively. Figure 7 presents the results of measurements made with our 1087-1 and 65132-2 sensors, and also with standard Vaisala sensors used in radiosondes RS80 and RS90. Readings of our sensors were compared with those of one of the standard Vaisala sensors, for various pressure values and the temperature $10^{o}$ and $20^{o}$C in the climate chamber in Lindenberg. This chamber is used for calibration of radiosonde sensors, and provides conditions with specified values of temperature, pressure, and humidity. Figure 8 presents the results of the comparison made for one of our sensors. The study confirmed that these sensors may be used for testing physical conditions in the cell and made it possible to trace out means for increase of the accuracy and reliability of these measurements.

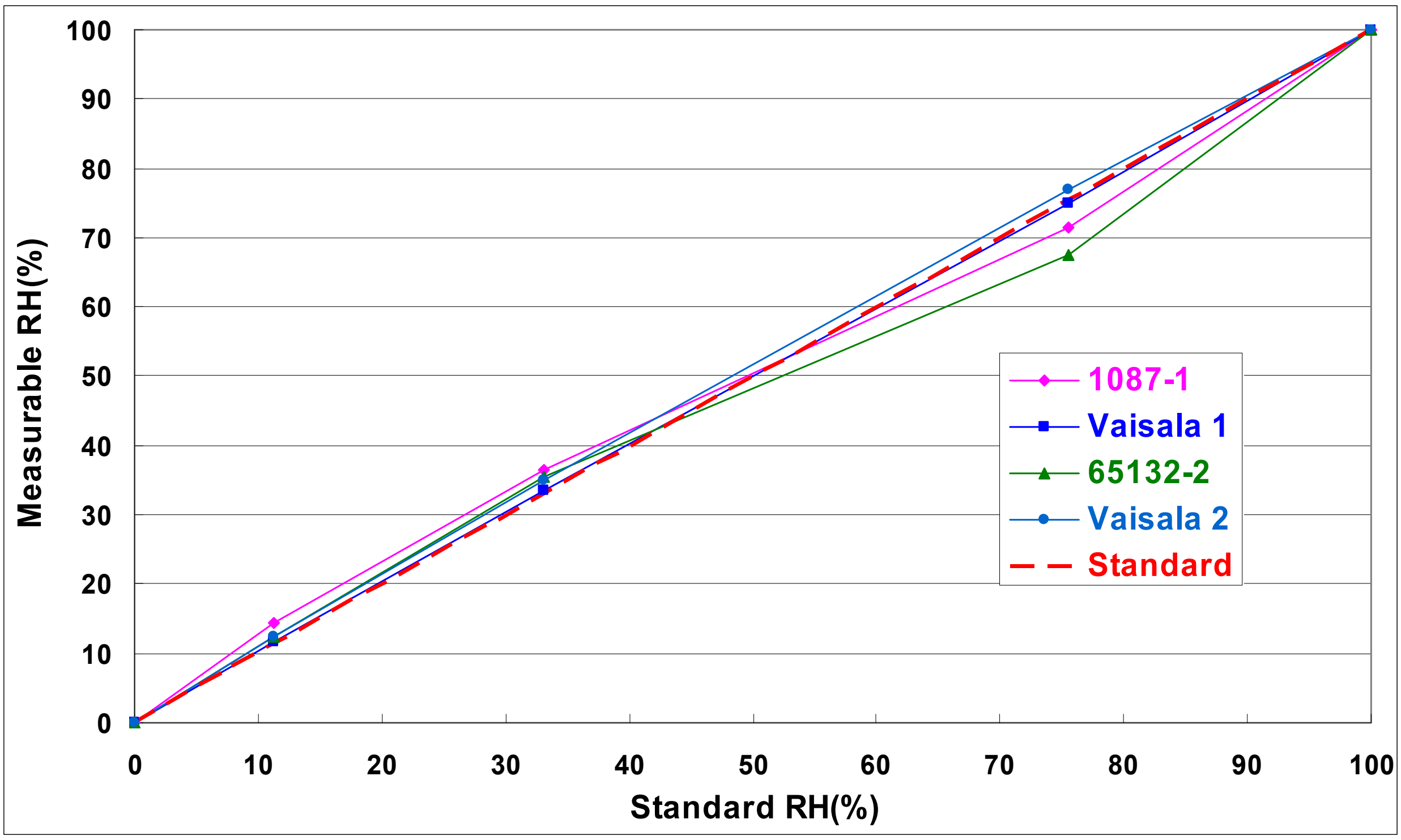

**Fig. 7**. Calibration of the sensors with the use of a standard saline generator of humidity.

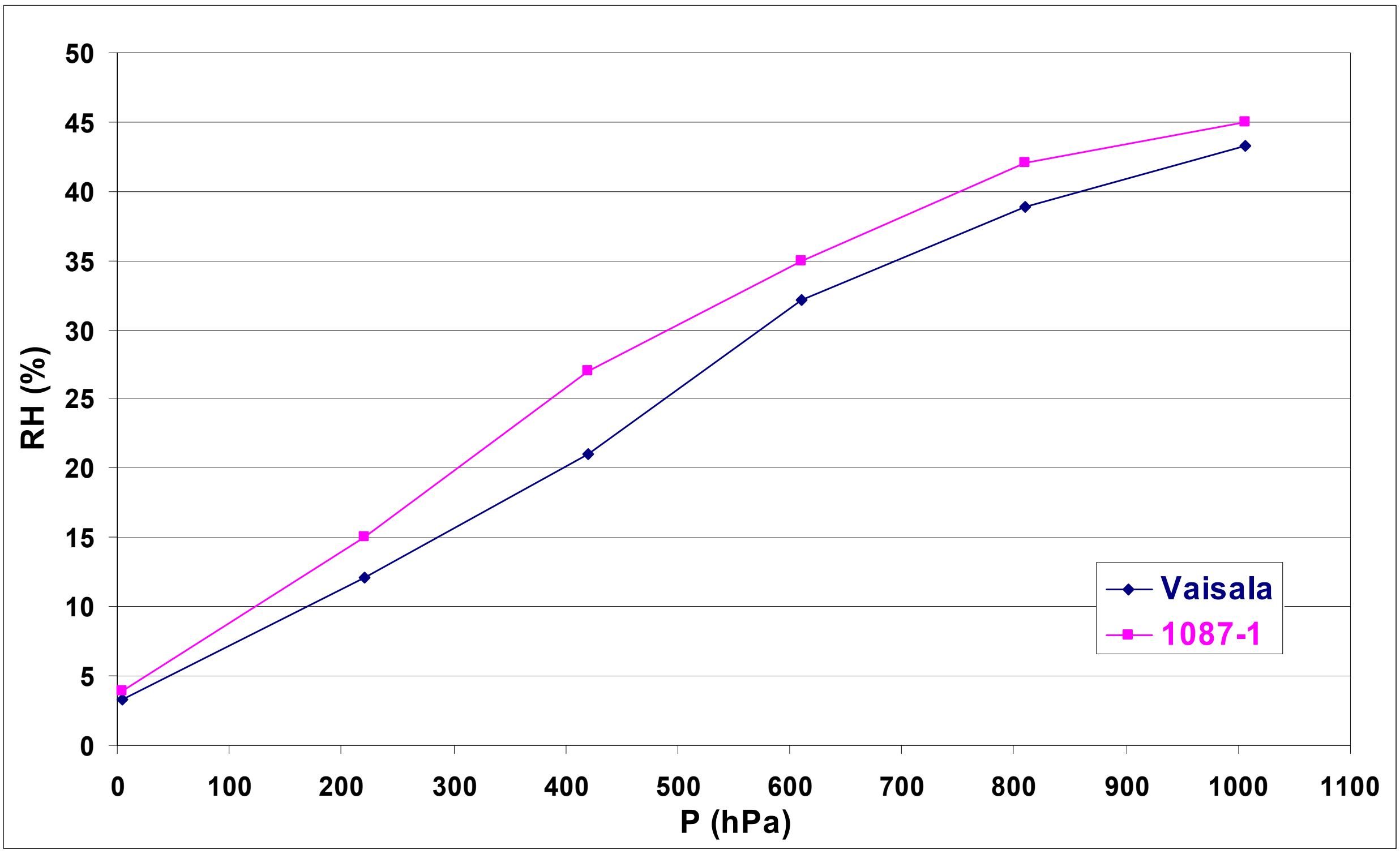

**Fig. 8**. A comparison between readings of the 1087-1 sensor and Vaisala sensor for various pressure values.

Another technique that makes it possible to determine the water vapor content in the cell is the optical method (implemented with the ASP-12 spectrograph optically connected with the VKM-100 cell), which uses the 6943.8 Å absorption line of water vapor. We applied this method in our study [19]. The advantage of this technique is that we determine the amount of water vapor directly along the line of view, which makes inhomogeneities along the path insignificant. However, the intensity of the 6943.8 Å line, used for this determination, may be inconsistent with the scale formed on the basis of saline solutions. Compared to the study [19], we revised the system of photoelectrical registration of the spectra, implementing the data input to the computer in the process of observations, which substantially increased the rate of data processing. Further on, we plan to combine the above techniques, so that the sensors continually test physical conditions in the cell, while the optical method provides means for testing the homogeneity of the medium and justifies the extension of the sensor measurements to the total volume of the cell. The intensity of the original line will be refined and corrected on the basis of readings of the sensors taken under the summertime conditions, when stable and uniform temperature conditions are established in the cell. This combined approach will make it possible to use a single initial source of calibration, both for the laboratory calibration of photometers, and for calibration of sensors used in atmosphere probing.

**Calibration of photometers.**

Currently, the optical method of determination of the atmospheric water vapor content is basically implemented with the use of solar and stellar photometers, in which spectral bands are discriminated with individual filters (Fig. 1) within the 930 nm $H_20$ absorption band and beyond it. The photometer should be optically interfaced with the cell: an optical unit (independent or optically conjugated with the photometer) should make it possible to observe the positions of the light spots on the main mirror of the cell, to discriminate the light beam that has passed a definite distance, and to fix its position on the entrance window of the photometer. The observation procedure consists in subsequent measurements of light that has passed through the cell different number of times, under fixed pressure and humidity, with a selected set of filters. The same procedure is then repeated for the evacuated cell. Since the intensity largely depends on the number of light passages, only observations obtained for the same number of passages may be compared. The comparison made for the filled and evacuated cell makes it possible to derive the absorption in the region of the water vapor band. When observations carried out at different time are compared, inaccuracies of the alignment, which are almost unavoidable, are removed by bringing into coincidence readings of filters placed beyond the absorption band. Figure 9 presents the obtained absorption as a function of the path length for the pressure of 1 atmosphere, for filters of the Pulkovo PZF-94 photometer.

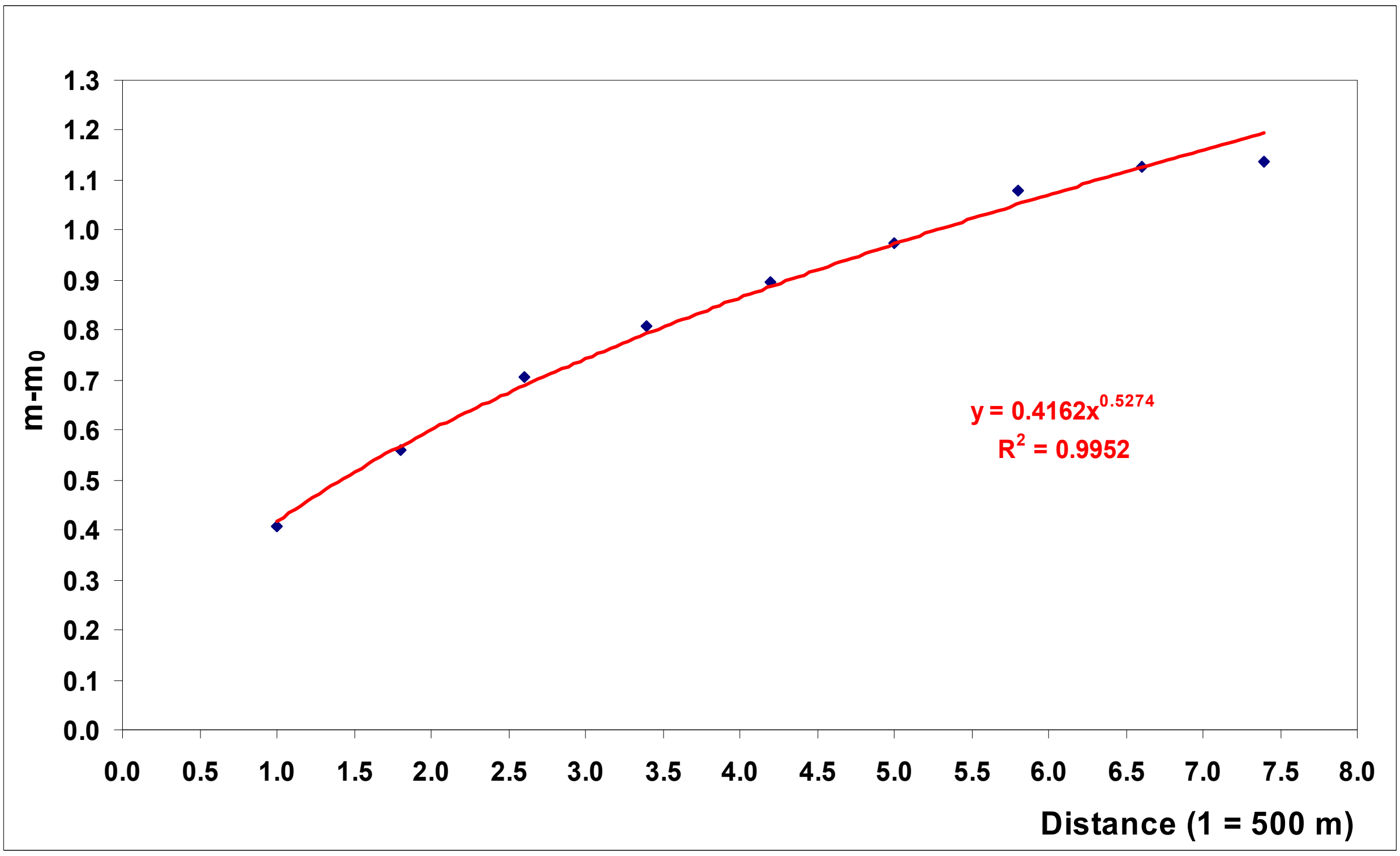

**Fig. 9**. The absorption by water vapor as a function of the length of the light path in the cell.

Here, the absorption is expressed in stellar magnitudes, the path length in the units of the minimum path length, equal to 500 m. Thereby, the Figure appears to be completely similar to the dependence of the atmospheric absorption on the air mass, where the unity air mass corresponds to the path length of 500 m. In the same way as the air mass may contain various amounts of water vapor, the path length of 500 m may also correspond to various amounts of water. The approximation of the observational data presented in the Figure by the expression (3) makes it possible to obtain the parameters **c** and **μ**. The former corresponds to the water vapor content along the unit path, i.e., 500 m. In our case, the water content varied from 0.54 to 4 cm of precipitated water along the line of view. Similar data, presented in Table 1, were also obtained for other conditions.

**Table 1**.

| | Set 1 | | Set 2 | |
|---|---|---|---|---|
| | Amount of water vapor along the line of view W (cmppw) | | | |
| Pressure (atm) | W(min) | W(max) | W(min) | W(max) |
| 1 | 0.037 | 0.306 | 0.539 | 4.417 |
| 0.9 | 0.040 | 0.330 | 0.526 | 4.313 |
| 0.8 | 0.043 | 0.351 | 0.512 | 4.201 |
| 0.7 | 0.043 | 0.355 | 0.708 | 5.805 |
| 0.6 | 0.044 | 0.361 | 0.638 | 5.233 |
| 0.5 | 0.032 | 0.265 | 0.598 | 4.907 |
| 0.4 | 0.041 | 0.340 | 0.520 | 4.263 |
| 0.3 | 0.027 | 0.223 | 0.477 | 3.908 |
| 0.2 | 0.034 | 0.283 | 0.361 | 2.958 |
| 0.1 | 0.018 | 0.146 | 0.219 | 1.800 |

Here, for various pressure values, the minimum and maximum amount of water vapor along the line of view is indicated, for which the measurements were carried out. The measurements were made with the Pulkovo PZF-94 (with both a photomultiplier and SPCM detector), and the Lindenberg stellar photometer.

Thereby, our study has demonstrated the performance efficiency of our complex. We have obtained experimental data on the absorption by water vapor for its content along the line of view from 0.02 to 5.8 cm of precipitated water, and also experimental calibration dependences for the indicated interval of the water vapor content. Further on, the total interval of water vapor content for conditions typical for on-sky observations should be modeled, the degree of reliability for the expression (3) should be determined, the limits for the variation of the obtained empirical parameters as a function of the variation of the water vapor content should be revealed, the transformation of the parameters with the variations of pressure should be studied, and the reliability of the transition from a uniform isothermal layer of air, studied in the laboratory, to the conditions of real inhomogeneous and non-isothermal atmosphere should be estimated.

For solar photometers, in order to compare directly the data obtained from laboratory calibrations with real atmospheric measurements, we suggest supplementing the above complex with the AZS-2 coelostat, which makes it possible to project the image of the Sun directly onto the entrance of the ASP-12 vacuum spectrograph. In this case, it will become possible to measure independently (from the 6943.8 Å water vapor absorption line) the real amount of water vapor in the atmosphere (along the line of view) and compare it with the results simultaneously obtained with the solar photometer, with the use of the parameters **c** and **μ** derived from laboratory measurements and radiosonde data.

## Conclusion

Here, we describe the basic elements of the laboratory complex for the calibration of photometers used in weather service to measure the water vapor content in the Earth atmosphere. The complex was developed within the framework of collaboration between Pulkovo

Observatory and Lindenberg Meteorological Observatory (Germany) and is primarily aimed at calibration and testing of Lindenberg's stellar and solar photometers. The problem is not only to obtain calibration dependences for particular instruments, but also to develop and compare different methods for construction of calibration dependences, based on direct calibration of photometers, on spectral laboratory functions of water vapor transmission, on calculation techniques that use spectroscopical databases for individual lines. In the course of this study, we also plan to refine the data on absorption capacity of water vapor, to establish possible impact of specific features of transmission curves of filters on empirical calibration dependences of photometers, to reveal the type of the influence of temperature and humidity on characteristics of instrumental systems of photometers. We hope that taking into account detailed parameters of the equipment and the use of new results concerning the absorption capacity of water vapor will make it possible to maintain the photometrical accuracy reached in observations (~0,5%) and to raise the accuracy of determination of the atmospheric water vapor content up to 1-2%.

The authors thank German Research Foundation (Deutsche Forschungsgemeinschaft, DFG) and the Russian Foundation for Basic Research (RFBR), which supported the given study by their grants (DFG- Project 436 RUS 113/76/0, DFG- Project 436 RUS 113/632/0-1, and RFBR grant 01-05-04000 NNIO-a).